# Cryogenic-temperature Grain-to-grain Epitaxial Growth of High-quality Ultrathin CoFe Layer on MgO Tunnel Barrier for High-performance Magnetic Tunnel Junctions

*Tomohiro Ichinose,[1,*] Tatsuya Yamamoto,[1] Takayuki Nozaki,[1] Kay Yakushiji,[1] Shingo Tamaru,[1] Shinji Yuasa[1]*

1. National Institute of Advanced Industrial Science and Technology (AIST), Research Center for Emerging Computing Technologies, Umezono 1-1-1, Tsukuba, 305-8568, Japan.

*Corresponding author
E-mail: tomohiro.ichinose@aist.go.jp

*Abstract*

One candidate for ultimate non-volatile memory with ultralow power consumption is magneto-resistive random-access memory (VC-MRAM). To develop VC-MRAM, it is important to fabricate high-performance magnetic tunnel junctions (MTJs), which require the epitaxial growth of an ultrathin ferromagnetic electrode on a crystalline tunnel barrier using a mass-manufacturing-compatible process. In this study, the grain-to-grain epitaxial growth of perpendicularly magnetized CoFe ultrathin films on polycrystalline MgO (001) was demonstrated using cryogenic-temperature sputtering on 300 mm Si wafers. Cryogenic-temperature sputtering at 100 K suppressed the island-like initial growth of CoFe on MgO without hampering epitaxy. Sub-nanometer-thick CoFe layers exhibited remarkable perpendicular magnetic anisotropy (PMA). An even larger PMA was obtained using an Fe-doped MgO (MgFeO) tunnel barrier owing to improved uniformity of the CoFe layer. A 0.8-nm-thick CoFe layer grown on MgFeO exhibited a magnetic damping constant as low as 0.008. The ultralow magnetic damping enables voltage-driven magnetization switching with a low write-error rate (WER) below $10^{-6}$ at a pulse duration of 0.3 ns, and WER on the order of $10^{-3}$ even for a relatively long pulse duration of 1.5 ns. These properties achieved using a mass-manufacturing deposition process can promote the development of VC-MRAM and other advanced spintronic devices based on MTJs.

*Introduction*

One promising candidate for energy efficient non-volatile memory (NVM) is magneto-resistive random-access memory (MRAM) devices. They are classified into several types in terms of data writing mechanisms, for example, toggling with Oersted fields, spin-transfer torque (STT), spin-orbit torque, and voltage-controlled magnetic anisotropy (VCMA). [1-5] In particular, VCMA-based MRAM is expected to be an ultimate NVM with ultralow power consumption and high write-endurance, which can replace current volatile cache memory devices. An integral part of these MRAMs is a magnetic tunnel junction (MTJ) with a high tunnel magnetoresistance (TMR) ratio, which comprises two ferromagnetic layers insulated with a crystalline tunneling barrier layer. [6,7] One of the ferromagnetic layers is magnetically fixed, while the other can be freely switched. The fixed layer employs a perpendicular synthetic antiferromagnetic (SAF) structure comprising antiferromagnetically coupled ferromagnetic layers with a nonmagnetic spacer layer. [8-11] In particular, hard SAF layers with high-temperature annealing tolerance have been obtained using Co/Pt superlattices coupled with a Ru or Ir spacer layer. [9-11] Because a flat and crystalline oriented buffer layer (e.g., (0001)-Ru) is required to fabricate Co/Pt superlattices, bottom fixed/top free type MTJs are desirable for MRAM applications. For the top-free layer, amorphous CoFeB-based multilayers, such as CoFeB/Mo/CoFeB, were fabricated on a MgO barrier layer because the MgO/CoFeB interface is known to simultaneously exhibit a high TMR ratio and interfacial perpendicular magnetic anisotropy (PMA). [12-17] High TMR ratio and PMA are required to obtain a sufficient readout margin and data retention time, respectively. In addition to these characteristics, it is important to reduce the magnetic damping in the free layer to decrease the switching current density in STT [1,18] and the write-error rate (WER) in voltage-driven magnetization switching. [19-22] However, reducing magnetic damping in a CoFeB-based free layer is difficult because a B-absorbing layer such as Mo or Ta, acting as extrinsic source of spin scattering, should be attached to the CoFeB layer to promote crystallization at MgO/CoFeB interfaces. [18] Furthermore, difficulty in thinning of the CoFeB-based free layer is problematic for voltage-driven magnetization switching because the coercivity changes due to the application of voltages tend to decrease with increasing free layer thickness.

Eliminating a B-absorbing layer may lead to substantially low magnetic damping, which can be achieved using a B-free crystalline ferromagnet surrounded by insulating layers, such as MgO. The use of a crystalline free layer will not only be advantageous for achieving lower magnetic damping but also enable the utilization of functionalities specific to crystalline ferromagnets. Examples include FePd, Mn-based ordered alloys with a large bulk PMA, [23,24] and Heusler alloys with half-metallic and/or spin-gapless band structures. [25,26] In addition, enhanced VCMA coefficients have been reported for single-crystal CoFe ultrathin films with diluted heavy metals. [27-30] The reason why those crystalline materials were not practically utilized was because thinning of crystalline magnetic films grown on a MgO barrier layer caused island-like growth due to low surface energy of MgO, which disperse and degrade the PMA in the top-free layer. [31] Therefore, suppression of island-like growth can boost the developments of crystalline ferromagnets for practical devices. Since island-like growth can originate from atomic migration during the formation of crystal grains, the suppression of atomic migration by lowering the growth temperature may be effective in suppressing island-like growth. Another possible method is to improve the wettability of the ferromagnet on the barrier surface through appropriate material selection. Improved wettability of the top-free layer was recently reported for the MgFeO/CoFeB interface. [32-34]

In this study, we developed MTJs with CoFe ultrathin films grown on an MgFeO barrier as the first example of a crystalline top-free layer fabricated at cryogenic temperature using a mass-production compatible sputtering system. The MTJs showed TMR ratios as high as 170% and PMA with effective anisotropy fields of 0.19 T after annealing at 673 K. In addition to the large TMR and PMA, which are comparable to those of conventional CoFeB-based MTJs, a low damping constant of 0.008 was obtained even in a 0.8-nm-thick ultrathin CoFe free layer due to the absence of a nonmagnetic B-absorbing layer. The use of the crystalline CoFe top-free layer resulted in a write-error reduction in voltage-driven magnetization switching. The application of crystalline materials enables the use of various functionalities to advance the development of next-generation MRAMs and numerous MTJ-based devices.

*Experimental Methods*

Typical MTJs in this work comprised buffer layers/Pt(0.25)/[Co(0.25)/Pt(0.15)]$_4$/Co(0.35)/Ir(0.48)/Co(0.85)/Mo(0.3)/Co$_{17}$Fe$_{53}$B$_{30}$(0.8)/MgO(1.7) or MgO(0.4)/Mg$_{40}$Fe$_{10}$O$_{50}$(1.3)/Co$_{50}$Fe$_{50}$($t_{CoFe}$)/MgO(1) (thicknesses in nm), which were deposited on $\phi$300 mm Si/SiO$_2$ wafers using a mass-production-compatible sputtering system (EXIM, Tokyo Electron Ltd.). $t_{CoFe}$ was 0.8 nm unless otherwise noted. A MgO/Mg$_{40}$Fe$_{10}$O$_{50}$ bilayer barrier was used in accordance with a previous study. [33] Hereafter, Mg$_{40}$Fe$_{10}$O and Co$_{50}$Fe$_{50}$ are simply denoted as MgFeO and CoFe, respectively. CoFe was deposited at $T_{CoFe}$ = 100 or 300 K to investigate the effect of deposition temperature, and Co/Pt multilayers in the SAF structure were deposited at 623 K. The remaining layers were then deposited at 300 K. The MTJs were post annealed ex-situ at $T_a$ = 673 K using a vacuum furnace with a base pressure of 8×10$^{-4}$ Pa under zero magnetic field. The multilayer films were patterned into circular pillars with diameters of approximately 90 nm using electron-beam lithography.

The structural properties of the MTJs were investigated using scanning transmission electron microscopy (STEM) analyses. A scanning transmission electron microscope equipped with energy-dispersive X-ray spectroscopy (EDX) and electron energy loss spectroscopy (EELS) was used to evaluate the elemental distributions in the MTJs. The magnetic properties of the MTJs were measured using vibrating sample magnetometry (VSM) and vector network analyzer ferromagnetic resonance (VNA-FMR). [33,35] The FMR spectra were measured under various perpendicular magnetic fields at a fixed microwave frequency. A probe station equipped with an electromagnet was used to investigate the magnetotransport properties of the MTJ devices including the TMR effect, differential conductance, and voltage-driven magnetization switching.

*Experimental Results*

Figs. 1(a) and 1(b) show cross-sectional dark-field STEM images of the MgO/CoFe MTJs with $T_{CoFe}$ = 300 and 100 K, respectively, after post-annealing at 673 K. The CoFe layer deposited at 300 K exhibited a rough and ambiguous interface compared to that deposited at 100 K, which indicates the island-like growth of CoFe ($T_{CoFe}$ = 300 K) due to the low surface energy of MgO. The island-like growth of CoFe was suppressed at $T_{CoFe}$ = 100 K, as shown in Fig. 1(b), indicating the effectiveness of cryogenic temperature deposition for the preparation of ultrathin polycrystalline metals on insulating materials with low surface energies. The suppression of the island-like growth of CoFe deposited at 100 K was also observed in MTJs with an MgFeO barrier, as shown in Fig. 1(c). The use of the MgFeO barrier further improved the uniformity of the CoFe layer, which can be attributed to the improved wettability at the MgFeO/CoFe interface owing to the presence of metallic Fe at the MgFeO surface. [34]

Additional STEM analyses of the MgFeO/CoFe MTJs ($T_{CoFe}$ = 100 K) were performed to investigate the effect of post-annealing on the crystal structure and elemental distribution (Fig. 2). Figs. 2(a) and 2(b) show bright-field STEM images of the MgFeO/CoFe MTJs before and after post-annealing at 673 K, respectively. As shown in Fig. 2(a), the as-deposited CoFe layer and the MgFeO barrier exhibited clear (001)-oriented lattice patterns, even though the CoFe layer was deposited at 100 K. The lattice patterns were continuous from the MgFeO barrier to the MgO cap, that is, the MgFeO/CoFe/MgO multilayers were grown in a grain-to-grain epitaxial manner. This result shows that thin film deposition at 100 K does not hamper the epitaxial growth of CoFe on MgFeO while reducing atomic migration during the deposition process. Fig. 2(b) shows that the layered structure was maintained even after the annealing at 673 K. Figs. 2(c) and 2(d) show the EDX and EELS profiles of the MgFeO/CoFe MTJs before and after post-annealing at 673 K, respectively; the origin of x-axis was set to the peak center of CoFe. From these analyses, no significant atomic interdiffusion was observed in the CoFe layer after the annealing at 673 K, demonstrating the high annealing tolerance of the crystalline CoFe-based MTJs with a MgFeO tunneling barrier and a MgO capping layer acting as a diffusion barrier.

Fig. 3 shows the magnetic hysteresis curves of the MTJ films with MgO/CoFe ($T_{CoFe}$ = 300 and 100 K) or MgFeO/CoFe ($T_{CoFe}$ = 100 K) annealed at 673 K. The full loop of the magnetization curves shown in Fig. 3(a) reveal sharp magnetization reversal and large interlayer exchange coupling in the perpendicularly magnetized SAF structure, which enables the measurement of minor loops. The minor loops shown in Fig. 3(b) reveal the magnetization process of the CoFe free layer. The round shape of the magnetization curve obtained for MgO/CoFe ($T_{CoFe}$ = 300 K) originates from the island-like growth of CoFe, as shown in the STEM image in Fig. 1(a). Following the suppression of island-like growth by cryogenic temperature deposition at 100 K, MgO/CoFe ($T_{CoFe}$ = 100 K) exhibited improved interfacial PMA with clear coercivity and large remanent magnetization. PMA properties were further improved by replacing the MgO barrier by a MgFeO barrier. The sharp magnetization reversal in MgFeO/CoFe ($T_{CoFe}$ = 100 K) is attributed to the improved CoFe layer uniformity shown in Fig. 1(c). Another possible reason for the improved PMA may be the formation of an Fe-rich region at the MgFeO/CoFe interface owing to the segregation of metallic Fe from MgFeO. [32,33]

The dynamic magnetic properties of MgO/CoFe and MgFeO/CoFe annealed at 673 K were analyzed using VNA-FMR to investigate the difference caused by changing barrier materials. We focused on the case of $T_{CoFe}$ = 100 K because the CoFe layer deposited at 300 K did not exhibit clear FMR spectra. Fig. 4(a) shows the FMR frequency ($f_{FMR}$) as a function of the perpendicular magnetic field ($\mu_0 H$), whose x-intercept corresponds to the effective anisotropy field ($\mu_0 H_{k,eff}$), and Fig. 4(b) shows the $f_{FMR}$ dependence of FMR spectral line width ($\mu_0 \Delta H$), whose linear relation can be expressed

as follows. [33,35]

$$\mu_0 \Delta H = \frac{2h}{g\mu_B} \alpha_{\text{tot}} f_{\text{FMR}} + \mu_0 \Delta H_0$$

where $h$ and $\mu_B$ are Planck's constant and the Bohr magneton, respectively. $g$ is the Lande factor, which can be estimated from the slope of $f_{\text{FMR}}$ on $\mu_0 H$. [33,35] $\alpha_{\text{tot}}$ is the total magnetic damping constant, which includes intrinsic damping and extrinsic spin relaxation such as that caused by the spin pumping effect. $\mu_0 \Delta H_0$ is the measure of spatial inhomogeneity in anisotropy fields. The obtained $\alpha_{\text{tot}}$ and $\mu_0 \Delta H_0$ are listed in Table 1 along with $\mu_0 H_{\text{k,eff}}$ estimated from Fig. 4(a). While the difference in $\mu_0 H_{\text{k,eff}}$ between MgO/CoFe and MgFeO/CoFe is rather small, there is a remarkable difference in the dependence of $\mu_0 \Delta H$ on $f_{\text{FMR}}$ as shown in Fig. 4(b). The use of the MgFeO barrier reduced both $\alpha_{\text{tot}}$ and $\mu_0 \Delta H_0$, which is attributed to the improved uniformity in crystalline orientation. Note that the $\alpha_{\text{tot}}$ in MgFeO/CoFe was as small as 0.008 even for the 0.8-nm-thick CoFe owing to the suppression of spin pumping to the neighboring layers.

The annealing effect on the dynamical magnetic properties was investigated by comparing the FMR data of the as-deposited MgFeO/CoFe to those of the 673 K-annealed one (Figs. 4(a) and 4(b)). While the annealed MgFeO/CoFe exhibited a $\mu_0 H_{\text{k,eff}}$ of 0.19 T, the as-deposited MgFeO/CoFe exhibited a negative $\mu_0 H_{\text{k,eff}}$, i.e., in-plane magnetic anisotropy. Moreover, larger $\mu_0 \Delta H_0$ and $\alpha_{\text{tot}}$ were observed in the as-deposited MgFeO/CoFe. Those results indicate that annealing at 673 K effectively improves interfacial PMA and its dispersion, although STEM analyses reveal that the layered structure of MgFeO/CoFe and the elemental distributions remain largely unchanged after annealing compared to their states before annealing. As interfacial PMA is sensitive to Fe-O bonding, the uniformity of the interfacial bonding may be improved by annealing.

Fig. 5(a) shows the TMR ratio in the MgFeO/CoFe MTJ ($T_{\text{CoFe}} = 100$ K, $T_a = 673$ K) defined as $(R-R_p)/R_p \times 100$ %, where $R$ ($R_p$) is junction resistance (at parallel magnetization configuration). Fig. 5(b) shows the bias voltage $V_{\text{bias}}$ dependence of the TMR ratio. A positive $V_{\text{bias}}$ is defined such that the electrons flow from top to bottom, as shown in the inset of Fig. 5(b). Thus, the positive (negative) region in Fig. 5(b) reflects the quality of the CoFeB/MgFeO (MgFeO/CoFe) interface. [36] The MgFeO/CoFe MTJ exhibited TMR ratio of 170% at $V_{bias}$ of 50 mV, along with sharp magnetization switching. A TMR ratio comparable to that of ultrathin CoFeB-based MTJs was obtained in perpendicularly magnetized MTJs using a crystalline CoFe top-free layer. The TMR ratio monotonically decreased with increasing $V_{bias}$ and became the half value at $V_{bias}$ of -0.7 and +1 V. The steep decrease in TMR ratio for the negative $V_{bias}$ indicates interfacial qualities such as flatness and crystal orientation at the MgFeO/CoFe interface may be inferior to those at the CoFeB/MgFeO interface. Fig. 5(b) shows the differential conductance ($di/dv$) curves measured in parallel and antiparallel magnetization configurations. Regardless of the magnetization configuration, the $di/dv$ curves were mostly parabolic, as observed in fundamental tunnel conduction, and did not exhibit clear fine structures reflecting the density of states. Fine structures can emerge by further improving the interfacial qualities of CoFe, which would also contribute to improve the TMR ratio. [36]

As we have shown above, improved magnetic properties such as $\mu_0 H_{k,eff}$ of 0.19 T, $\alpha_{tot}$ of 0.008, and TMR ratio of 170% were demonstrated in 0.8-nm-thick crystalline CoFe grown at 100 K on the MgFeO barrier. Achieving a low $\alpha_{tot}$ is crucial for a free layer in MRAM applications. In particular, the free layer in the voltage-controlled MRAMs should be as thin as possible while minimizing the enhancement of $\alpha_{tot}$, meaning the ultrathin CoFe layer with low damping can be a promising candidate for voltage-controlled MRAMs. Thus, we examined the performance of the CoFe free layer from the viewpoint of the VCMA effect and demonstrated voltage-driven magnetization switching. In this study, an ultrathin Ir dust layer was used to enhance the VCMA effect in CoFe. [27-30]

Figs. 6(a) and 6(b) show the $V_{bias}$ dependence of the TMR curves of the MgFeO/Ir(0 or 0.06 nm)/CoFe MTJs. The polarity of $V_{bias}$ is the same as that in Fig. 5, i.e., a positive (negative) $V_{bias}$ induces charge accumulation (depletion) at the MgFeO/CoFe interface, which decreases (increases) coercive fields $\mu_0H_c$ of the CoFe free layer. The variation of $\mu_0H_c$ is plotted in Fig. 6(c) as a function of $V_{bias}$. The slopes in Fig. 6(c) ($\mu_0 \Delta H_c/\Delta V$) was -20 and -10 mT V$^{-1}$ for MTJs with and without Ir, respectively, indicating Ir insertion enhances the VCMA effect in polycrystalline CoFe, as observed in epitaxial CoFe MTJs. PMA can be tuned by varying the CoFe thickness $t_{CoFe}$ for voltage-driven magnetization switching. In fact, MgFeO/Ir(0.06 nm)/CoFe(0.95 nm) exhibited a $\mu_0H_c$ of 48 mT with $\mu_0 \Delta H_c/\Delta V$ of -29 mT V$^{-1}$, which indicated that the $\mu_0H_c$ can be eliminated by applying a $V_{bias}$ of 1.7 V. The voltage-driven magnetization switching due to the elimination of the $\mu_0H_c$ was demonstrated under an in-plane $\mu_0H$ of 53 mT as shown in Fig. 6(d). Fig. 6(d) shows the WER in voltage-driven magnetization switching defined as $1-p_{sw}$, where $p_{sw}$ is the magnetization switching probability for applying the voltage pulse. The WER oscillates as a function of pulse duration, which is a signature of VCMA-driven magnetization switching. [37] The first half-period of magnetization precession was observed at a pulse duration of 0.3 ns, for which the lowest WER, less than 10$^{-6}$, was achieved. Furthermore, a low WER of the order of 10$^{-3}$ was maintained even in the third local minimum at a pulse duration of 1.5 ns, in contrast to the results of previous studies reporting WER on the order of 10$^{-1}$ under the long (> 1 ns) pulse duration. [38] The reduction in WER can be attributed to the low magnetic damping of CoFe. [19-22]

***Conclusion***

We fabricated top-free-type MTJs with polycrystalline CoFe using cryogenic temperature deposition and a MgFeO barrier. These MTJs exhibited a TMR ratio of 170% and clear PMA with a $\mu_0H_{k,eff}$ of 0.19 T, in addition to the low magnetic damping of 0.008, owing to the negligible spin pumping in the top-free structure surrounded by insulating materials. The high performances, resulting from the use of cryogenic temperature deposition and the MgFeO barrier, can be attributed to the improved interface flatness, sharpness, and uniform crystalline orientation, as evidenced by STEM analyses. The improvements in interfacial qualities due to cryogenic temperature deposition and the MgFeO barrier can also be applied to other functional crystalline materials, which paves the way for advances in future MTJ applications, such as MRAM devices. A voltage-driven writing operation was demonstrated using the MgFeO/CoFe-based MTJs. The MgFeO/CoFe MTJ exhibited a low WER on the order of $10^{-3}$ even for a long pulse duration of 1.5 ns as well as a WER less than $10^{-6}$ at pulse duration of 0.3 ns. The low WER can be attributed to the low magnetic damping of the CoFe ultrathin films.

*Acknowledgments*

This study was partly based on the results obtained from a project, JPNP16007, subsidized by the New Energy and Industrial Technology Development Organization (NEDO). The TEM analyses were performed with the support of the Foundation for Promotion of Material Science and Technology of Japan (MST).

*Conflict of Interest*

The authors declare no conflict of interest.

Table 1. Magnetic parameters in top-free CoFe extracted from ferromagnetic resonance.

| Sample | $T_a$ [K] | $\mu_0 H_{k,\text{eff}}$ [T] | $\alpha_{\text{tot}}$ | $\mu_0 \Delta H_0$ [mT] |
|---|---|---|---|---|
| MgO/CoFe/MgO | 673 | 0.16 | 0.055 | 38 |
| MgFeO/CoFe/MgO | 673 | 0.19 | 0.008 | 19 |
|  | As deposited | -0.37 | 0.076 | 38 |

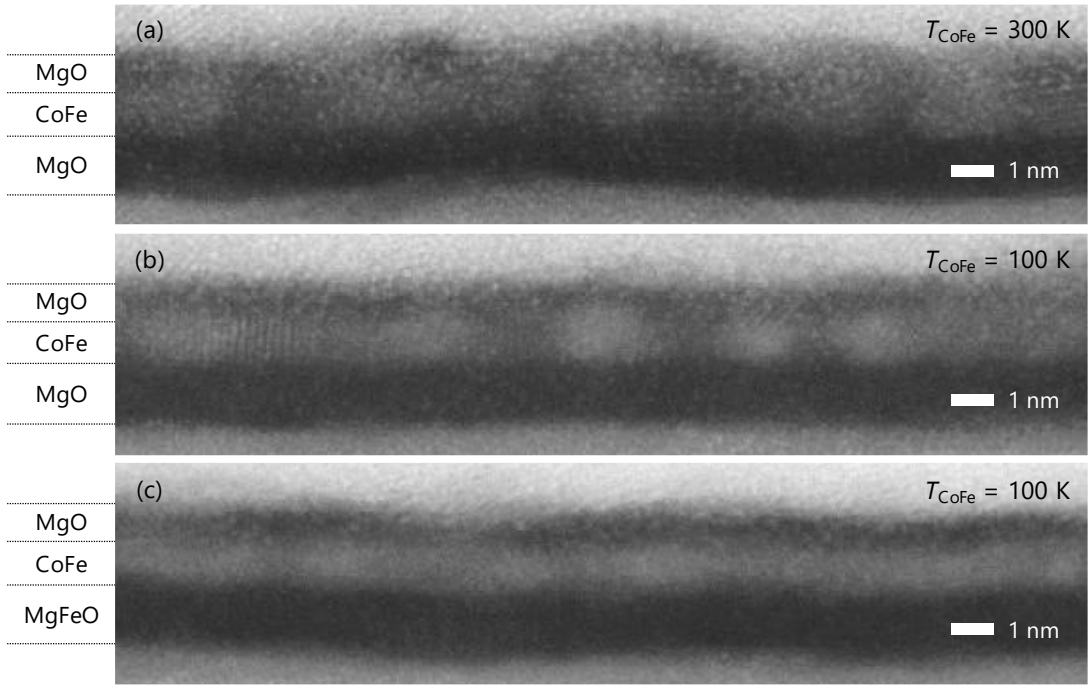

Fig. 1. Dark-field STEM images for (a) MgO/CoFe ($T_{\text{CoFe}}$ = 300 K), (b) MgO/CoFe ($T_{\text{CoFe}}$ = 100 K), and (c) MgFeO/CoFe ($T_{\text{CoFe}}$ = 100 K) annealed at 673 K.

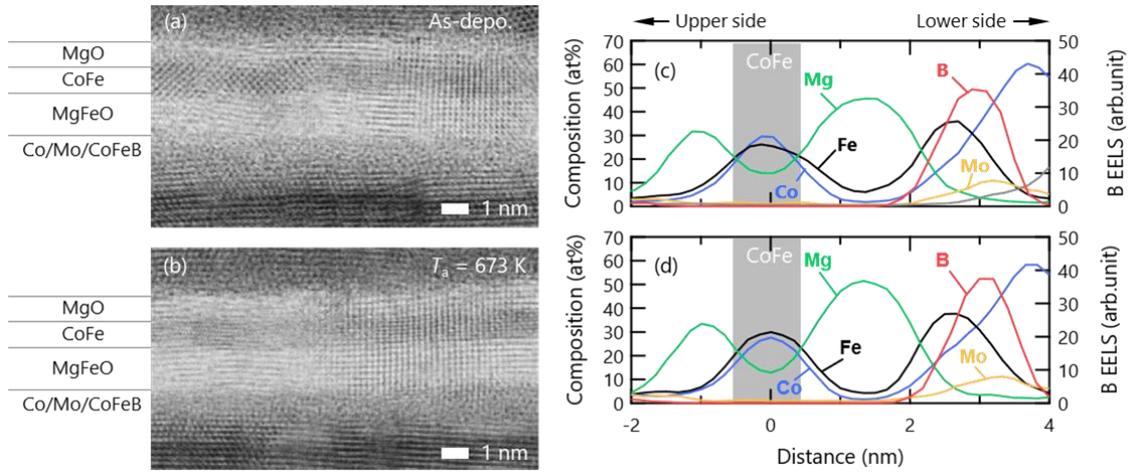

Fig. 2. Bright-field STEM images for MgFeO/CoFe ($T_{CoFe}$ = 100 K) (a) before and (b) after annealing at 673 K. (c) and (d) are elemental distributions corresponding to (a) and (b), respectively. Elemental distributions were measured with EDX except for B. B was detected by EELS.

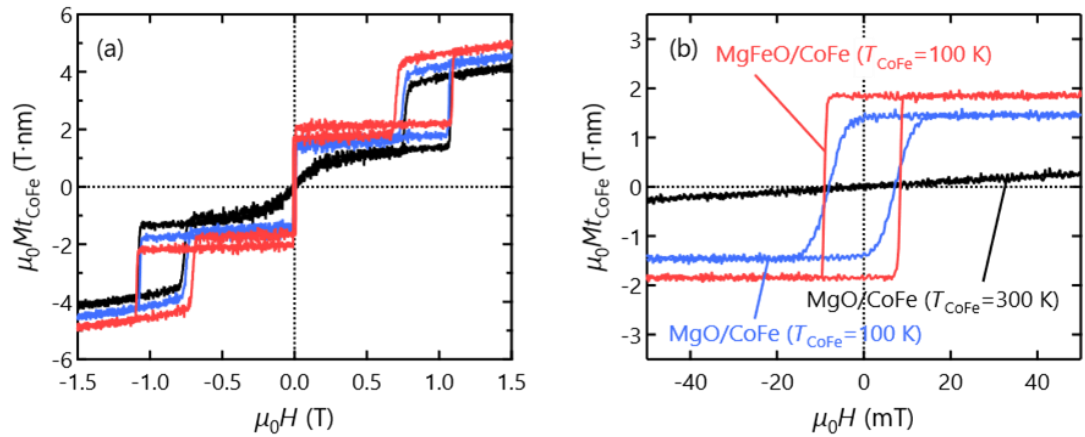

Fig. 3. (a) Full and (b) minor magnetization curves of MTJs with MgO/CoFe ($T_{CoFe}$ = 300 K), MgO/CoFe ($T_{CoFe}$ = 100 K), and MgFeO/CoFe ($T_{CoFe}$ = 100 K) annealed at 673 K.

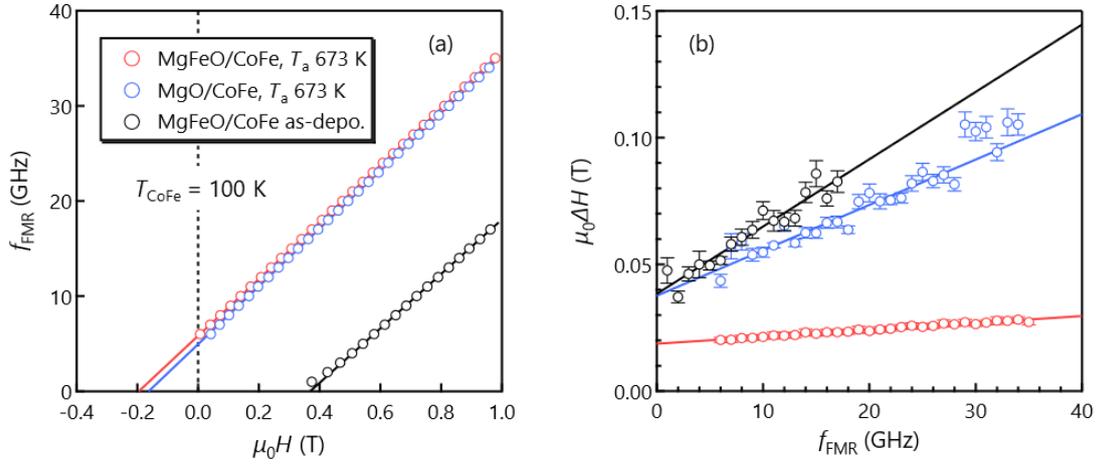

Fig. 4. (a) FMR frequency $f_{FMR}$ as a function of perpendicular magnetic field $\mu_0 H$, whose x-intercept corresponds to the effective anisotropy field. (b) $f_{FMR}$ dependence of spectral line width $\mu_0 \Delta H$, in which the y-intercept and slope indicate spatial inhomogeneity in the anisotropy field and total magnetic damping, respectively.

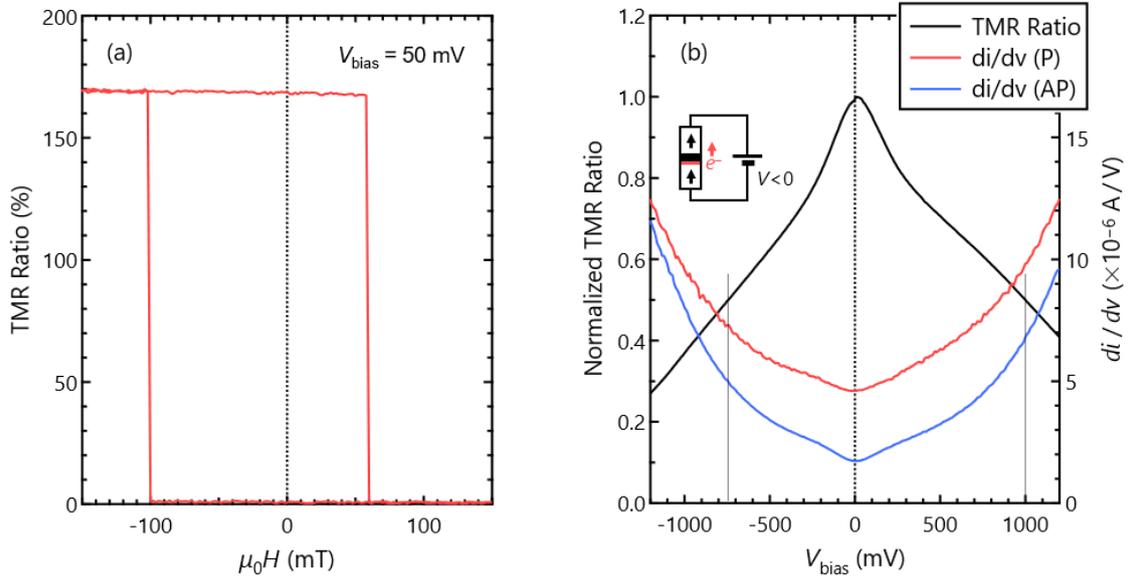

Fig. 5. (a) The TMR curve and (b) the dependence of the TMR ratio and differential conductance $di/dv$ on bias voltage $V_{bias}$ for a MgFeO/CoFe MTJ ($T_{CoFe}$ = 100 K) annealed at 673 K.

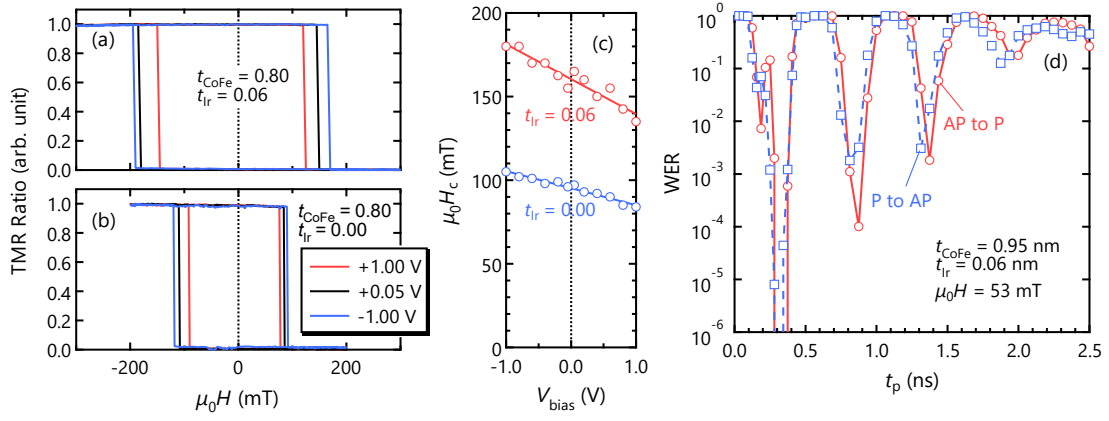

Fig. 6. TMR curves in MgFeO/CoFe MTJs (a) with and (b) without Ir(0.06 nm) insertion for various $V_{bias}$ values. (c) $V_{bias}$ dependence of coercive field $\mu_0 H_c$. (d) WER for voltage-driven magnetization switching in the MgFeO/Ir/CoFe MTJ under an in-plane magnetic field $\mu_0 H$ of 53 mT. P and AP indicate parallel and antiparallel magnetization configurations, respectively.